\documentclass[aps,a4paper,superscriptaddress,twocolumn,showpacs]{revtex4}
\usepackage{graphics,epsfig}
\usepackage{setspace}

\newcommand{\be}{\begin{equation}}
\newcommand{\ee}{\end{equation}}
\newcommand{\bea}{\begin{eqnarray}}
\newcommand{\eea}{\end{eqnarray}}
\newcommand{\ba}{\begin{eqnarray*}}
\newcommand{\ea}{\end{eqnarray*}}

\begin{document}
\title{Gold Clusters Sliding on Graphite: \\a Possible Quartz Crystal Microbalance Experiment?}

\author{S. Pisov}
\affiliation{International Center for Theoretical Physics (ICTP),
P.O.Box 586, I-34014 Trieste, Italy} \affiliation{University of
Sofia, Department of Atomic Physics, 5 James Bourchier Blvd., 1164
Sofia, Bulgaria}
\author{E. Tosatti}
\affiliation{International School for Advanced Studies (SISSA)
and Democritos,
Via Beirut 2-4, I-34014 Trieste, Italy} \affiliation{International
Center for Theoretical Physics (ICTP), P.O.Box 586, I-34014
Trieste, Italy}
\author{U. Tartaglino}
\affiliation{International School for Advanced Studies (SISSA)
and Democritos,
Via Beirut 2-4, I-34014 Trieste, Italy}
\author{A. Vanossi}
\affiliation{CNR-INFM National Research Center S3 and Department
of Physics, \\University of Modena and Reggio Emilia, Via Campi
213/A, 41100 Modena, Italy}

\date{December 4, 2006}
\begin{abstract}

A large measured 2D diffusion coefficient of gold nanoclusters on
graphite has been known experimentally and theoretically for about
a decade. When subjected to a lateral force, these clusters should
slide with an amount of friction that could be measured. We
examine the hypothetical possibility to measure by Quartz Crystal
Microbalance (QCM) the phononic sliding friction of gold clusters
in the size range around 250 atoms on a graphite substrate between
$300$ and $600$ K. Assuming the validity of Einstein's relations
of ordinary Brownian motion and making use of the experimentally
available activated behavior of the diffusion coefficients, we can
predict the sliding friction and slip times as a function of
temperature. It is found that a prototypical deposited gold
cluster could yield slip times in the standard measurable size of
$10^{-9} sec$ for temperatures around $450-500$ K, or $200$ C.
Since gold nanoclusters may also melt around these temperatures,
QCM would offer the additional chance to observe this phenomenon
through a frictional change.\\
Preprint of: {\it J. Phys.: Condens. Matter} {\bf 19} (2007) 305015.\\
Link: http://stacks.iop.org/0953-8984/19/305015

\end{abstract}

\maketitle

\section{Introduction}
Understanding the diffusion mechanisms and the frictional
properties of aggregates of atoms or molecules of nanometric size
({\it nanoclusters}) on surfaces is important from both
fundamental and technological view points. Growth of new materials
with tailored features, as, e.g., a structure controlled down to
the nanometer scale, is one of the active research fields in
physical science. Different experimental techniques can be used to
build nanostructured systems; however the requisites of control
(in terms of characterization and flexibility) and efficiency (in
terms of quantity of matter obtained per second) are generally
incompatible. The main advantage of the cluster-deposition
technique is that one can carefully control the building block
(i.e., the cluster) and characterize the formation processes. The
behavior of these deposited nano-objects is often distinctive
(mostly due to their large surface-to-volume ratio), being
qualitatively different from those of their constituent parts and
from those of bulk material. In particular, they may present
properties that vary dramatically with size.

Under different circumstances, various individual mechanisms
(single-atom-like processes) can be responsible for cluster
motion, such as evaporation and/or condensation, diffusion of
particles along the cluster edge, and motion of misfit
dislocations. Differently, one of the remarkable experimental
observations of the last decade is concerned with gliding-like dynamics
of compact solid gold clusters as a whole (concerted jumps with conservation
of size and shape).

After room temperature deposition, gold clusters, comprising
typically 250 atoms (radius of $\sim 10$ \AA), have been repeatedly
observed to diffuse on Highly Ordered Pyrolytic Graphite (HOPG)
surfaces with surprisingly
large, thermally activated diffusion coefficients in the range
$10^{-7}-10^{-5}$ cm$^2$/sec already at room temperature
\cite{bardotti96}; a similar behavior was reported for Sb$_{2300}$
clusters. A large mobility was recently observed also for
activated surface diffusion of close-packed hexagonal clusters
Ir$_7$ and Ir$_{19}$ \cite{Wang}. The detailed atomic mechanisms
for diffusion of gold clusters, on which we focus here, was
studied theoretically and simulated by Molecular Dynamics (MD) by
Luedtke {\it et al.} \cite{landman99}, who pointed out the
coexistence of short and long jumps, the latter assimilating the
process to a L\'{e}vy flight. Later work by Lewis {\it et al.}
\cite{lewis2000} and by Maruyama \cite{maruyama04} further
explored the cluster diffusion mechanisms.

The observable diffusion of sizeable clusters as a whole raises an
interesting question in the context of nanofriction. If one could
manage to apply a sufficient lateral driving force to the
clusters, they would drift under its action. By the
fluctuation-dissipation theorem, the lateral cluster drift
mobility should then be related (leaving Levy flights aside for
the time being \cite{footnote}) to diffusion through Einstein
relation. The measurable diffusivity of deposited clusters could
thus lead to a measurable frictional dissipation in experimental
apparatuses such as a Quartz Crystal Microbalance (QCM)
\cite{krim, mistura}. This is precisely the issue we propose to
explore synthetically in this paper.

Specifically, we wish to examine a hypothetical, yet very practical, QCM
experimental case where sliding friction could be measured for
gold clusters adsorbed on graphite, for sizes around $250$ atoms
and temperatures between room temperature and about $600$ K. As
it will turn out, the experimentally observed temperature
dependence of the diffusion coefficients (obeying an
Arrhenius-type activated law) predicts, through Einstein's
relation, easily measurable QCM slip times at temperatures of the
order of $450$--$500$ K, possibly even lower, which anyway appear
well within the reach of existing QCM setups.

An additional interesting effect is that small gold clusters will
tend to ``premelt'' already at much lower temperatures than the
bulk $T_m$ = 1336 K. Buffat and Borel \cite{buffat} analyzed free
gold clusters for premelting. Their data actually suggest melting
not too far from room temperature for sizes around Au$_{250}$ or
thereabout. In addition one may expect the solid to acquire some
additional stability against the liquid when adsorbed on the flat
graphite substrate. So the actual temperature where the adsorbed
$Au$ cluster melts into a (partly) wetting $Au$ droplet is
difficult to predict and will have to await experimental scrutiny.
Our point however is precisely that a sudden change of QCM sliding
friction as T increases should be generically observable, and will
signal that the cluster has melted.

\section{Modelling}

The simplest theoretical model for cluster diffusion could be a
model of jump over simple monatomic steps. A single jump
mechanism predicts diffusion coefficients given by
\begin{equation}
\label{arrhenius:eqn} D(T) = \frac{\nu a^2}{4} \exp \left(
-\frac{E_d}{k_B T}\right)
\end{equation}
where $a$ is the jump step, $\nu$ is the attempt frequency, and
$E_d$ is the energy barrier to be overcome. However, even if $\nu$
is taken as large as the Debye frequency ($\nu \approx 10^{12}
s^{-1}$) , with $a \sim 0.3$ $nm$, the prefactor is $10^{-3}$
cm$^2$/sec \cite{siclen95}, many orders of magnitude smaller than
experimental value of $10^3$ cm$^2$/sec \cite{bardotti96}.

Luedtke {\it et al.} \cite{landman99} simulated gold cluster
diffusion numerically using a many-body embedded atom (EAM) potential for
the interaction among the gold atoms, and a two-body Lennard-Jones
(LJ) potential for the interaction between gold and carbon atoms.
The LJ potential parameters were $\epsilon_{Au-C} = 0.01273$ eV
and $\sigma_{Au-C} = 2.9943$ \AA. The $Au-C$ LJ potential was
given different parameters $\epsilon_{Au-C} = 0.022$ eV and
$\sigma_{Au-C} = 2.74$ \AA\ in a later simulation by Lewis {\it et
al.} \cite{lewis2000}. Both works uncovered a cluster diffusion
mechanism that has a very interesting nature. There are long ``sticking''
periods of short-range local dwelling without too much diffusion of
the cluster. Every now and then, there appear rare but important events where the
cluster ``slips'' -- it actually glides -- over relatively long distances.
The sticking plausibly corresponds to trapped cluster configurations,
occasionally abandoned thanks to large and rare fluctuations. During
one such fluctuation the two crystalline surfaces in contact, the gold
cluster and the graphite plane, behave as hard incommensurate sliders
\cite{Dienwiebel}, and thus slip for a while essentially freely.

This reasoning is supported by existing numerical results based on
molecular dynamics simulations. Many-particle clusters that are
incommensurate with the substrate were shown \cite{jensen96} to
exhibit very rapid diffusion, with their paths akin to a Brownian
motion induced by the internal vibrations of the clusters and/or
the vibrations of the substrate. The behavior of an incommensurate
object moving ``as a whole'' is in striking contrast with other
diffusion mechanisms, especially observed for clusters epitaxially
oriented on the surface, where the cluster motion results from a
combination of single-atom processes (e.g., evaporation,
condensation, edge diffusion, etc.). The latter mechanisms, giving
rise to relatively low diffusion coefficients ($D \sim 10^{-17}$
cm$^2$/sec), are likely not significant in cases where the
mismatch between cluster and substrate is large and/or their
mutual interaction is weak. In this view, a gold cluster adsorbed
with a (111) face onto a graphite substrate forms an ideal
mismatched (incommensurate) system with two very stiff mating
surfaces that interact only weakly, thus representing a desirable
sliding situation.

\begin{figure}
\epsfig{file=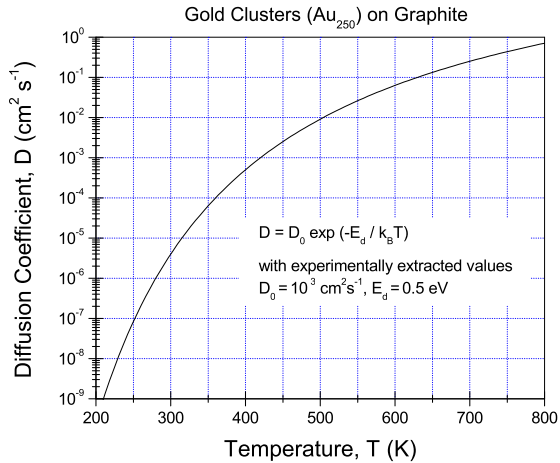,width=8.5cm,angle=0,clip=}
\caption{\label{activation:fig}
Temperature dependence of the diffusion coefficient of Au$_{250}$
on graphite. The values of $D_0$ and $E_d$ are chosen to fit the
experimental data by Bardotti {\it et al.} \cite{bardotti96} with
an Arrhenius-type law.
}
\end{figure}

As shown by the simulation work by Lewis {\it et al.}
\cite{lewis2000}, this stick-slip type dynamics still leads --
despite its large difference from the simple jump model over
monatomic barriers -- to an activated diffusion coefficient of the
form \ref{arrhenius:eqn} above, although now with different
coefficients from monatomic diffusion. That work however also
clarifies the quantitative inadequacy of the LJ interaction model,
which for a 250 atom cluster leads to a prefactor of about
$2\times 10^{-3}$ cm$^2$/sec and an effective barrier of $0.17$
eV, against experimental values of $10^3$ cm$^2$/sec and $0.5$ eV
\cite{bardotti96}. The Arrhenius activated behavior fit to these
experimental data is shown in Fig.~\ref{activation:fig}. The
discrepancy between simulation and experiment was highlighted by
Maruyama \cite{maruyama04}.

The source of discrepancy between simulations and experiment is
probably related to a large error in the adsorption energy. An
experimental estimate for adsorption energy of Au on graphite is
$E_a = 0.64$ eV
\cite{anton00}. {\it ab-initio} calculations suggested even larger
values for the adsorption energy of about $0.9$ eV
\cite{jensen04}. On the other hand the LJ model potentials imply a
much smaller adsorption energies $E_a \sim 0.15$ eV. With this
smaller adsorbtion energy, a cluster could experience a smaller
effective energy barrier when trying to disentangle itself from
the graphite substrate. Thus the adsorption energy error, at first
sight irrelevant, might we suspect be at the origin of the large
discrepancy of effective barriers in cluster diffusion ($0.5$ eV
against $0.17$ eV in Lewis's LJ modeling, and probably even
smaller in Luedtke's).

While it will be important to fix these problems in future
simulations, we can for the time being content ourselves with this
qualitative understanding, and simply discuss cluster diffusion
as a regular Brownian diffusion, although with a rather unusual
activation mechanism.

\section{Estimate of the slip time through Einstein's relation}

When an adsorbate island or a cluster is forced to slide on a substrate, the
dissipation of energy due to kinetic friction can be characterized
by a specific tribological quantity, the {\it slip time} $\tau$,
defined as the time taken by the cluster speed, initially
set to be nonzero, to drop to $1/e$ of its original value. This is a
relaxation time associated with the cluster-momentum fluctuation, and
is connected to the interfacial friction coefficient $\eta$ through
the relation $\eta = \rho /\tau$, where $\rho$ is the mass per unit
area of the cluster. Thus, defining $\rho = m_{Au} N / A$ ($N$
denoting here the number of gold atoms of the cluster in direct contact
with the graphite substrate over an area $A$), the interfacial friction
coefficient can be rewritten as

\begin{equation} \label{eta:eqn}
\eta = m_{Au} N  / A \tau.
\end{equation}

Our assumption that cluster diffusion enjoys the properties of
Brownian motion implies the applicability of the
fluctuation-dissipation theorem, which in turn causes the force
free diffusion and the friction under a sliding force to be connected through
Einstein's relation

\begin{equation} \label{einstein:eqn}
D \eta A = k_B T.
\end{equation}

By substituting expression (\ref{eta:eqn}) for the interfacial
friction coefficient in (\ref{einstein:eqn}), we obtain for the
slip time

\begin{equation} \label{sliptime:eqn}
\tau = D m_{Au} N / k_B T.
\end{equation}

Through this relation, assuming approximately $N=50$ atoms
directly touching the graphite substrate and, from the
experimental data of Bardotti {\it et al.} \cite{bardotti96}, a
diffusion coefficient $D \sim 10^{-5}$ cm$^2$/sec for Au$_{250}$
nanoclusters at $300$ K, we get a rough estimate of the room
temperature slip time $\tau \sim 10^{-12}$ sec. While this value
is at least two orders of magnitude too small to be observed by
QCM, the situation, as discussed below, can change drastically
upon heating.

\section{A possible QCM experiment}

The problem of how to measure interfacial friction in a
quantitative manner has remained long unsolved, while knowledge of
this property would provide important information relevant to a
wide variety of problems. Sliding friction of adsorbed films are
very effectively probed by QCM. In a QCM experiment, a substrate
is laterally oscillated with typical frequencies $f \sim 15$ MHz
and amplitudes of $a \sim 100$ \AA. This exerts on a $N$-atom
adsorbate island to a typical inertial force $F = N (2\pi)^2 M f^2
a$, which when sufficiently large can depin the island and make it
slide. The sliding frictional dissipation lowers the oscillator's
quality factor $Q$, and the result is conventionally measured by
the slip time $ \tau = d(Q^{-1}) / df$. As was said earlier, the
interfacial friction coefficient may then be derived from
Eq.~\ref{eta:eqn}. Sliding friction vanishes in both limits $\tau$
= 0 (film locked to the substrate), and $\tau = \infty$
(superfluid film), and typical measurable slip times are in the
order of $10^{-8}-10^{-9}$ sec \cite{persson_book}.

\begin{figure}
\epsfig{file=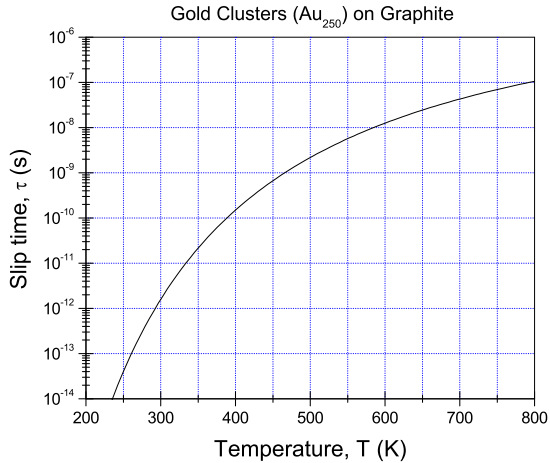,width=8.5cm,angle=0,clip=}
\caption{\label{sliptime:fig}
Predicted temperature dependence of the slip time of a Au$_{250}$ on graphite.
$\tau$ is evaluated considering an interfacial contact area of $N=50$
Au atoms.
}
\end{figure}

For low speed, hard sliders on hard flat substrates far from their
melting points, there is no wear, and all frictional dissipation occurs
via phonon creation (in metals, also creation of electron
hole pairs). We concentrate here on phononic dissipation, whose
origin stems from the oscillatory potential energy
variation associated with lateral cluster displacement, the
so-called ``corrugation''. A recent analysis of solid xenon
monolayers sliding over different substrates suggested for example
that the viscous friction coefficient $\eta$ increases quadratically
with corrugation\cite{krim05}

\begin{equation} \label{corrugation:eqn}
\eta = \eta_{subs} + bU^2_0
\end{equation}

\noindent where $\eta_{subs}$ is the dissipation arising from both
phononic and electronic friction, $b$ is a coefficient which
depends weakly from the substrate surface, and $U_0$ is the
potential corrugation amplitude between adsorbate layer and
substrate, controlling the phononic dissipation. Experimental upper bounds for
$\eta_{subs}$ are $0.08$ $ns^{-1}$ (for Xe/Cu(111), Xe/Ni(111),
and Xe/graphite).

We propose here a possible QCM experiment to study the frictional
dissipation of gold nanoclusters sliding on a graphite substrate
heated above room temperature. Raising  temperature will bring
about QCM-measurable slip times. Working at temperature even as
high as $\sim 600$ K does not affect the structural properties of
graphite, and should be entirely possible. Assuming for the moment
the Au clusters to remain solid, we can simply extrapolate at
higher temperatures the diffusion and slip times as done in
Fig.~\ref{sliptime:fig}. The predicted cluster slip time should
for a 250 atom cluster reach QCM measurable values of $10^{-9}$
sec between $T = 450$ and $500$ K, where the corresponding
diffusion coefficient $D \sim 10^{-2}$ cm$^2$/sec.

\section{Au clusters melting and nanofriction}
\begin{figure}
\epsfig{file=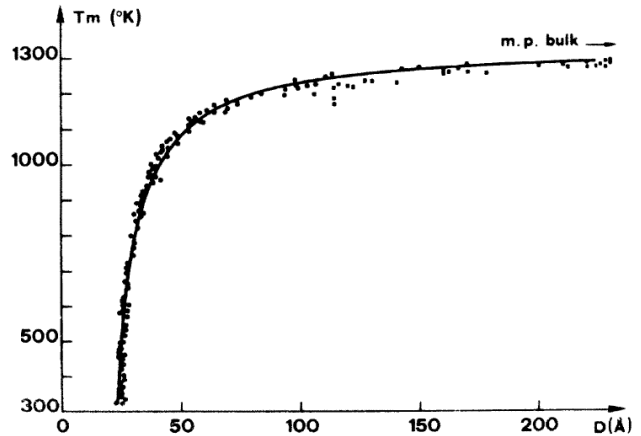,width=8.5cm,angle=0}
\caption{\label{melting:fig}
Experimental values of the melting point temperature of free gold
clusters as a function of their diameter. From Ref.~\cite{buffat}.
}
\end{figure}
As was recalled in the introduction, small clusters of radius $R$ have
a premelting temperature $T_m(R)$ well below the bulk melting
temperature of the same material. Figure~\ref{melting:fig}
reproduces the original data by Buffat and Borel \cite{buffat},
showing that a small gold cluster of diameter $\sim 20$ \AA\ will
generally melt at room temperature or below. Premelting of gold clusters
has also been extensively simulated, notably by Ercolessi {\it et al.}.\cite{ercolessi}

The premelting temperature data for free gold clusters are well
fit by the Gibbsian formula \cite{tartaglino05}
\begin{equation} \label{melting:eqn}
T_m(R) \approx T_m(\infty) \left( 1 - \frac{2}{\rho_l L R}
\Delta\gamma \right),
\end{equation}
with $ T_m(\infty) = 1336$ K,  and $\Delta \gamma =
\gamma_{SV}-(\gamma_{SL}+\gamma_{LV}) \approx 600$ mJ/m$^2$.
Here, $\rho_l$ is the density of the liquid, $L$ is the latent
heat of melting per unit mass, and the $\gamma$'s denote the free
energies of the three solid-vapor (SV), solid-liquid (SL), and
liquid-vapor (LV) interfaces.

Even though a deposited cluster might generally be more stable than
a free one due to the stabilizing effect of a hard substrate, it will
still premelt very readily upon heating. At melting, one should
expect the nanocluster friction to change very significantly. In fact
the nanocluster will turn into a droplet, with some characteristic
wetting angle, and obviously losing its original rigidity. The droplet
dissipation should be dominated by viscosity, with a damping
whose magnitude is at this stage difficult to anticipate. The jump
between rigid cluster sliding and viscous droplet dissipation should
nonetheless be a strong feature, readily observable by QCM, taking
place below or near 600 K, depending on size, for clusters below
20 \AA in diameter.

In summary, we predict that QCM measurement of small gold clusters
on a graphite substrate should lead to observable inertial sliding
friction with measurable slip times at temperatures of a couple hundred
centigrades. The relatively sudden melting of deposited clusters should also
become measurable in the form of a sudden jump of slip time upon
heating, at temperatures that should depend very strongly on cluster size.

{\it Acknowledgments} - S.P. is grateful to ICTP's Condensed
Matter Group for sponsoring the visit that led to this work.
Partial support by MIUR COFIN 2004023199, by INFM/FIRB RBAU017S8R
and MIUR FIRB RBAU01LX5H is acknowledged. A.V. thanks the
financial support by PRRIITT (Regione Emilia Romagna), Net-Lab
``Surfaces \& Coatings for Advanced Mechanics and Nanomechanics''
(SUP\&RMAN). Discussions with R. Guerra are gratefully
acknowledged.

\end{document}